\documentclass[%
 reprint,
 amsmath,amssymb,
 aps,
]{revtex4-2}

\usepackage{graphicx}
\usepackage{dcolumn}
\usepackage{bm}
\usepackage{mathptmx}
\usepackage{etoolbox}
\usepackage{hyperref}
\usepackage{booktabs}
\usepackage{physics}
\usepackage{color}

\begin{document}

\preprint{APS/123-QED}

\title{Enhancement of Chiral-Induced Spin Selectivity via Circularly Polarized Light}

\author{Wei Liu}
\affiliation{Department of Chemistry, School of Science, Westlake University, Hangzhou 310024 Zhejiang, China}
\affiliation{Institute of Natural Sciences, Westlake Institute for Advanced Study, Hangzhou 310024 Zhejiang, China}
\author{Jingqi Chen}%
\affiliation{Department of Chemistry, School of Science, Westlake University, Hangzhou 310024 Zhejiang, China}
\affiliation{Institute of Natural Sciences, Westlake Institute for Advanced Study, Hangzhou 310024 Zhejiang, China}

\author{Wenjie Dou}
 \email{douwenjie@westlake.edu.cn}

\affiliation{Department of Chemistry, School of Science, Westlake University, Hangzhou 310024 Zhejiang, China}
\affiliation{Department of Physics, School of Science, Westlake University, Hangzhou 310024 Zhejiang, China}
\affiliation{Institute of Natural Sciences, Westlake Institute for Advanced Study, Hangzhou 310024 Zhejiang, China}

\date{\today}

\begin{abstract}
The notion of chiral-induced spin selectivity (CISS) has attracted intensive research interest recently. However, the practical applications of the CISS effects face challenges due to relatively low spin polarization. In this Letter, we propose a non-perturbative theory illustrating how circularly polarized (CP) light enhances CISS effects through strong light-matter interactions. We introduce a Floquet electronic friction model to study the nonadiabatic dynamics and spin transport through a chiral molecule in a molecule junction subjected to external driving. Our results show that the interplay of the nonadiabatic effects and light-matter interactions can significantly ($>90\%$) enhance electron spin polarization under CP light. Our predictions can be very useful in experiments for using CP light to control spin current in chiral molecular junctions. 
\end{abstract}

\maketitle

\textit{Introduction.}—Chirality characterizes parity-symmetry breaking where a molecule cannot be superposed on its mirror image in chemistry and biology~\cite{quack1989structure,siegel1998homochiral,barron2021symmetry}. Chiral organics have recently been reported to exhibit a topological feature~\cite{liu2021chirality}, in which the electronic orbital and momentum are locked together, to rationalize the intriguing spin selectivity in DNA-type molecules~\cite{naaman2019chiral,evers2022theory}, known as CISS~\cite{ray1999asymmetric}. 
The CISS effects were initially reported in 1999 by Naaman and collaborators\cite{ray1999asymmetric}. 
Since then, a number of additional experimental~\cite{gohler2011spin,xie2011spin,ben2014local,kiran2016helicenes,eckshtain2016cold,mondal2016photospintronics,alpern2016unconventional,alam2017spin,abendroth2017analyzing,kumar2017chirality,aragones2017measuring,varade2018bacteriorhodopsin,santos2018chirality,gazzotti2018spin,shapira2018unconventional,ghosh2019controlling} and theoretical~\cite{brandbyge2002density,nikolic2005decoherence,paulsson2008unified,yeganeh2009chiral,areshkin2010electron,guo2012spin,gutierrez2012spin,guo2014spin,wu2015spin,chang2015nonequilibrium,medina2015continuum,marmolejo2017proximity,michaeli2019origin,matityahu2016spin,varela2016effective,pan2016spin,matityahu2017spin,schmaltz2017effect,diaz2018effective,diaz2018thermal,palsgaard2018efficient,thoss2018perspective,maslyuk2018enhanced,nurenberg2019evaluation,kosov2018telegraph,rudge2016distribution,papior2017improvements,yang2019spin,holder2020consequences,liu2021chirality,batge2021nonequilibrium,wolf2022unusual,honeychurch2023quantum,lorentzen2023quantum,rudge2023current,adhikari2023interplay,wan2023anomalous,subotnik2023chiral,gersten2013induced,du2020vibration,wu2021electronic,teh2022spin} studies have been conducted. A future industrial applications of CISS is spin-based chip design, which should enable electronic devices
to reach atomic scale~\cite{kim2014spin,brandt2017added}. Although we can observe up to $85\%$ spin polarization in experiments using the contact magnetic atomic force microscope (AFM) setup~\cite{naaman2020chiral}, spin selection occurs prior to electron passage through the chiral molecule, and this initial spin selection is primarily due to the strong SOC of the metal substrate~\cite{gersten2013induced}. The SOC alone induced by electron motion within chiral molecules~\cite{wolf2022unusual} is too small to give rise to the observed spin polarization~\cite{zhang2020chiral}. How to enhance CISS has attracted extensive research.

Recent studies indicate that electron-vibration coupling enhances the CISS effect, leading to an increase in the spin polarization in double-helix DNA~\cite{du2020vibration}. When considering the coupling between atomic nuclei and electron spins, calculations reveal strong spin selectivity near the conical intersection, even when the SOC is small, with spin selectivity reaching up to $100\%$~\cite{wu2021electronic}.
In the presence of a non-zero current, Berry curvature can induce nuclear spin separation and electron spin polarization~\cite{teh2022spin}. However, the universal theory on how to enhance CISS is still lacking.

In physics, chirality usually refers to the spin–momentum locking of particles such as Weyl fermions~\cite{armitage2018weyl,yan2017topological} and CP light. Chiral enantiomers exhibit opposite chiroptical activity when coupling to light~\cite{barron1986symmetry}. Due to the dissymmetric interaction between CP light and chiral molecules, there is a selectivity in absorption or emission of left-handed versus right-handed CP light for chiral molecular systems~\cite{greenfield2021pathways, liu2024anomalous}. Therefore, inspired by the close connection between chiral molecules and light-matter interactions, we sought to explore whether CP light could enhance the CISS, contributing to the ongoing advancements in spintronics~\cite{naaman2019chiral,shang2022emerging}.

In this Letter, we present a non-perturbative theory to illustrate how circularly polarized (CP) light enhances CISS through strong light-matter interactions. Our results indicate that when chiral molecules interact with CP light, there is a significant ($>90\%$) enhancement of electron spin polarization. The prediction from our newly developed theory highlights the effects of using CP light to control spin current in a chiral molecular junction, which can be potentially verified in CISS experiments.

\begin{center}
\begin{figure}
    \centering
    \includegraphics[width=.98\textwidth]{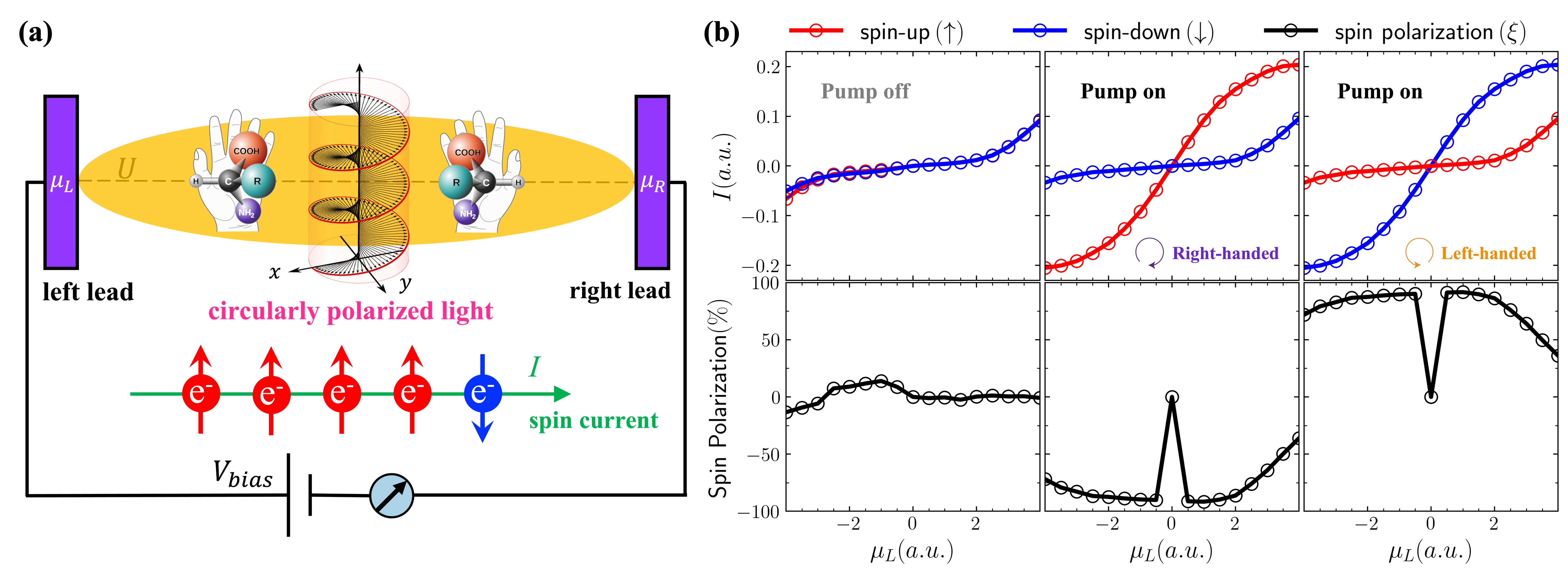}
    \caption{(a) A schematic picture of our chiral molecular junction model interacted with circularly polarized light. $\mu_{\text{L}}$ and $\mu_{\text{R}}$ are chemical potentials of the left and right leads, and the source-drain voltage $V_{\text{sd}} = \mu_{\text{L}} - \mu_{\text{R}}$ controls the voltage bias $V_{\text{bias}}$. (b) Steady-state spin current $I^{\uparrow/\downarrow}$ in Eq.~\ref{eqn:iupdown_m} and spin polarization $\xi$ in Eq.~\ref{eqn:sp} for $h_{\text{s}}$ (pump off) in Eq.~\ref{eqn:hs_soc}, $h_{\text{s}}^{\text{[R]}}$ (pump on and right-handed) in Eq.~\ref{eqn:hs_right} and $h_{\text{s}}^{\text{[L]}}$ (pump on and left-handed) in Eq.~\ref{eqn:hs_left}. For spin-up and spin-down electrons, we calculate the average spin current using $10^4$ nuclear motion trajectories with random force $\delta F_\mu^{\text{F}}$, respectively. Parameters: $\Delta=3$, $A=B=1$, $\Omega=1$, $kT = 0.5$, $\lambda_x =0$, $\lambda_y =0.8$, $\widetilde{\Gamma} = 1$, $\mu_{\text{L}} = -\mu_{\text{R}}$ and Floquet level $N=5$.}
    \label{fig:schematic}
\end{figure}
\end{center}

\textit{General Hamiltonian.}—As shown in the schematic diagram [Fig.~\ref{fig:schematic} (a)], we start from a minimal model for CISS with two spatial orbitals and spin-orbit couplings between them~\cite{gohler2011spin,teh2022spin}. These orbitals couple to left and right leads whose voltages are $\mu_{\text{L}}$ and $\mu_{\text{R}}$, respectively. In addition, the Hamiltonian depends on two nuclear degrees of freedom (x and y are considered, uniformly represented as $\mathbf R$). The total model Hamiltonian $\hat H_{\text{tot}}$ consists of the kinetic energy and another three parts: the system Hamiltonian $\hat H_{\text{s}}$, the bath Hamiltonian $\hat H_{\text{b}}$ composed of left and right leads, and the system-bath coupling Hamiltonian $\hat H_{\text{sb}}$,
\begin{eqnarray}
&\hat{H}_{\text{tot}} = \frac{\hat{P}^2}{2M}+\hat H_{\text{s}}  + \hat H_{\text{b}} + \hat H_{\text{sb}},   \\
&\hat H_{\text{s}}  = \sum_{ij} [h_\text{s}]_{ij} (\mathbf R, t) \hat d_i^\dagger \hat d_j + U(\mathbf R), \\
&\hat H_{\text{b}} = \sum_{k\zeta } \epsilon_{k\zeta  } \hat c_{k\zeta}^\dagger \hat c_{k\zeta },  \\
&\hat H_{\text{sb}}  =  \sum_{\zeta k,i} V_{\zeta k,i}  (   \hat c_{k\zeta}^\dagger \hat d_i + \hat d^\dagger_i  \hat c_{k\zeta}  ).\label{eqn:sb}
\end{eqnarray}
In system Hamiltonian ($\hat H_{\text{s}}$), the operators $\hat d_i^\dagger$ and $\hat d_i$ denote create and annihilate an electron in the $i$-th spin orbital of the subsystem respectively. $U(\mathbf R)$ represents the electrostatic potential between nuclei. As for the Hamiltonian ($\hat H_{\text{b}}$), the operators $\hat c_{k\zeta}^\dagger$ and $\hat c_{k\zeta}$ create and annihilate electrons in the $k$-th spin orbital of the lead $\zeta$, where the energy of the orbital is denoted by $\epsilon_{k\zeta}$. Lastly, in the interaction Hamiltonian ($\hat H_{\text{sb}}$), the tunneling element $V_{\zeta k,i}$ indicates the interaction between subsystem spin orbital $i$ and lead spin orbital $k\zeta$. In Eq.~\ref{eqn:sb}, the Condon approximation has been utilized, implying that $V_{\zeta k,i}$ is independent of $\mathbf R$.

For the system Hamiltonian, we consider chiral molecules interacting with a CP light beam, which involves periodic driving from light-matter interactions. For right-handed/clockwise CP light, the system Hamiltonian $h_{\text{s}}^{\text{[R]}}$ writes
\begin{eqnarray}
\begin{aligned}
    &h_\text{s}^{\text{[R]}}(\mathbf{R},t) = \\
    &\begin{bmatrix}
     x+ \Delta & Ax\cos(\Omega t)-\text{i}By\sin(\Omega t) \\
     Ax\cos(\Omega t)+\text{i}By\sin(\Omega t) & -x - \Delta
    \end{bmatrix}.
\end{aligned}
\label{eqn:hs_right}
\end{eqnarray}
While for left-handed/counterclockwise CP light, the system Hamiltonian $h_s^{\text{[L]}}$ writes
\begin{eqnarray}
\begin{aligned}
    &h_\text{s}^{\text{[L]}}(\mathbf{R},t) = \\
    &\begin{bmatrix}
     x+ \Delta & Ax\cos(\Omega t)+\text{i}By\sin(\Omega t) \\
     Ax\cos(\Omega t)-\text{i}By\sin(\Omega t) & -x - \Delta
    \end{bmatrix}.
\end{aligned}
\label{eqn:hs_left}
\end{eqnarray}
Finally, for no polarized light, the system Hamiltonian $h_\text{s}$ writes 
\begin{eqnarray}
\begin{aligned}
    h_\text{s}(\mathbf{R},t) = 
    \begin{bmatrix}
     x+ \Delta & Ax-\text{i}By \\
     Ax+\text{i}By & -x - \Delta
    \end{bmatrix}.
\end{aligned}
\label{eqn:hs_soc}
\end{eqnarray}
Here, $\Delta$ represents the energy gap between two spatial orbitals, parameter $A$ denotes the shifts of diabatic PES and $B$ denotes the strength of spin-orbit couplings. Moreover, $\Omega$ denotes the frequency of the periodic driving. 


The shifted parabola model is extensively employed to simulate electron transfer and excitation energy transfer processes~\cite{nitzan2006chemical}. Note that Eq.~\ref{eqn:hs_right}, Eq.~\ref{eqn:hs_left} and Eq.~\ref{eqn:hs_soc} are all effective Hamiltonians, where positive $B$ represents spin-up state and negative $B$ represents spin-down state.

As commonly used in literature~\cite{malhado2012photoisomerization}, the bare  potential $U(\mathbf R)$ is set to be a two dimensional parabolas: 
\begin{equation}
U(\mathbf R) = \frac{1}{2}(x - \lambda_x{x})^2 + \frac{1}{2}(y - \lambda_y{y})^2.
\end{equation}
Even though the scalar potential $U(\mathbf R)$ does not couple to spin state directly, the linear terms $\lambda_x$ and $\lambda_y$ in $U(\mathbf R)$ can change the spin current and spin polarization dramatically. 
At this point, we have introduced our model for spin transport in a chiral molecule with optical control. 
\onecolumngrid
\twocolumngrid

\textit{Floquet Nonadiabatic Dynamics.}—We adapt our newly developed Floquet electronic friction model to study nonadiabatic spin transport dynamics~\cite{mosallanejad2023floquet, chen2023floquet}. In particular, all electronic DoFs (including spin-orbit couplings) as well as periodic driving are being incorporated into frictional force and random force on nuclear motion. As a result, one can simply run Langevin equation of motion for the nuclei, which is given by 
\begin{equation}
  M_\mu \ddot R_\mu = F_\mu^{\text{F}} - \sum_\nu \gamma_{\mu\nu}^{\text{F}} \dot R_\nu +  \delta F_\mu^{\text{F}}.
\label{eqn:langevin_m}
\end{equation}
Here, $F_\mu^{\text{F}}$, $\gamma_{\mu\nu}^{\text{F}}$, and $\delta F_\mu^{\text{F}}$ are mean force, friction tensor, and random force respectively. We use the superscript ${\text{F}}$ to denote their Floquet counterparts. To evaluate $F_\mu^{\text{F}}$, $\gamma_{\mu\nu}^{\text{F}}$, and $\delta F_\mu^{\text{F}}$, we need to represent $h_{\text{s}}^{\text{[R]}}$ and $h_{\text{s}}^{\text{[L]}}$ in Floquet levels.
Here, we provide an explicit formula for each block of the extensive matrix, which is in the Fourier space:
\begin{eqnarray}
h_{\text{s}}^{\text{F}} = \sum_{n = -N}^{N} [h_{\text{s}}^{[\text{R}]/[\text{L}]}]^{(n)}\hat{L}_n + \hat{N}\otimes\hat{I}_n \hbar \omega,
\label{eqn:hsF}
\end{eqnarray}
\begin{eqnarray}
[h_{\text{s}}^{[\text{R}]/[\text{L}]}]^{(n)} = \frac{1}{T}\int_{0}^{T}h_{\text{s}}^{[\text{R}]/[\text{L}]}(\mathbf R,t)e^{-\text{i}n\Omega 
 t}dt,
\label{eqn:hsn}
\end{eqnarray}
where $\hat{N}$ is the number operator, $\hat{I}_n$ is the identity matrix for the $n$-th order and $T$ is the period, equal to $2\pi/\Omega$. $\hat{L}_n$ is the n-th ladder operator which turns the vector-like Fourier expansion into a matrix-like representation. Specifically, Eq.~\ref{eqn:hsF} represents the decomposition of $h_{\text{s}}^{[\text{R}]/[\text{L}]}(\mathbf R,t)$ into different energy components rotating at different frequencies in the Fourier space. The term $[h_{\text{s}}^{[\text{R}]/[\text{L}]}]^{(n)}$ denotes the decomposition of $h_{\text{s}}^{[\text{R}]/[\text{L}]}(\mathbf R,t)$ at the frequency of $n\Omega$. The first term of Eq.~\ref{eqn:hsF} is the weighted sum of each Floquet potential, while the second term is the tensor product of $\hat{N}$ and $\hat{I}_n$. This term can be added to the Hamiltonian because $\hat{N}$ commutes with the Hamiltonian. Eq.~\ref{eqn:hsF} allows us to understand the time evolution of the system at different rotating frequencies, and further investigate the physical properties of the system.

The explicit form of the frictional force and random force are computed directly in terms of Green's functions. The frictional force is given by
\begin{equation}
\gamma_{\mu\nu}^{\text{F}} = - \frac{\hbar}{2\pi (2N+1)}\\
\int_{-\infty}^{+\infty}d\epsilon \text{Tr}\left \{ \partial_{\mu} h_{\text{s}}^{\text{F}} \partial_{\epsilon} G_{\text{r}}^{\text{F}} \partial_{\nu} h_{\text{s}}^{\text{F}} G_{<}^{\text{F}} \right \} + \text{H.c.}.\\
\end{equation}
Please refer to Supplemental Material (SM) A for details.

\textit{Spin Current.}—The spin current $I^{\uparrow/\downarrow}$ is obtained by averaging over the nuclear DOFs with respect to local current~\cite{teh2022spin}:
\begin{eqnarray}
I^{\uparrow/\downarrow} = \int d\mathbf R d\mathbf P I_{\text{loc}}^{\text{F}}(\mathbf R) \rho^{\uparrow/\downarrow}(\mathbf R, \mathbf P). 
\label{eqn:iupdown_m}
\end{eqnarray}
The local current $I_{\text{loc}}^{\text{F}}$ flowing from the left lead through the molecule to the right lead can be evaluated by the Landauer formula~\cite{haug2008quantum, meir1992landauer},
\begin{eqnarray}
\begin{aligned}
I_{\text{loc}}^{\text{F}} &=\\
&\frac{e}{2\pi\hbar(2N+1)} \int_{-\infty}^{+\infty}d\epsilon \text{Tr} \left \{ T^{\text{F}}(\epsilon)[f_{\text{L}}^{\text{F}}(\epsilon)-f_{\text{R}}^{\text{F}}(\epsilon)] \right \}. \\
\end{aligned}
\label{eqn:iloc_m}
\end{eqnarray}
In the above equation, $f_{\text{L}}^{\text{F}}(\epsilon)$ and $f_{\text{R}}^{\text{F}}(\epsilon)$ represent the Floquet Fermi-Dirac distribution functions that correspond to the left and right leads, respectively. Additionally, $T^{\text{F}}(\epsilon)$ denotes the Floquet transmission probability, which can be deconstructed using the Green's functions.
\begin{eqnarray}
T^{\text{F}}(\epsilon) = \Gamma_{\text{L}}^{\text{F}} G^{\text{F}}_r(\epsilon) \Gamma_{\text{R}}^{\text{F}} G_{\text{a}}^{\text{F}}(\epsilon),
\label{eqn:tran}
\end{eqnarray}
where
\begin{eqnarray}
\begin{aligned}
\Gamma_{\text{L}}^{\text{F}} = \Gamma_{\text{L}} \otimes \hat{I}_n, \quad
\Gamma_{\text{R}}^{\text{F}} = \Gamma_{\text{R}} \otimes \hat{I}_n.
\end{aligned}
\end{eqnarray}

In this project, we examine the distinct interaction between a molecule and two leads where solely orbital 1 ($x+\Delta$) couples to the left lead and only orbital 2 ($-x-\Delta$) couples to the right lead. This arrangement is demonstrated in the subsequent definition of the $\Gamma$ matrices:
\begin{eqnarray}
 \Gamma_{\text{L}} = 
    \begin{pmatrix}
      \widetilde{\Gamma} & 0 \\
      0  &  0
    \end{pmatrix},\quad
 \Gamma_{\text{R}} = 
    \begin{pmatrix}
       0& 0 \\
      0  &  \widetilde{\Gamma}
    \end{pmatrix}.
\end{eqnarray}

Finally, we define spin polarization $\xi$ as follows, 
\begin{eqnarray}
\xi = \frac{I^{\downarrow} - I^{\uparrow}}{I^{\downarrow} + I^{\uparrow}} \times 100 \%.
\label{eqn:sp}
\end{eqnarray}
Here, $I^{\uparrow/\downarrow}$ denotes spin-up or spin-down current. 
See SM B for details of computing $I^{\uparrow/\downarrow}$.

\begin{figure}[htbp]
\includegraphics[width=.48\textwidth]{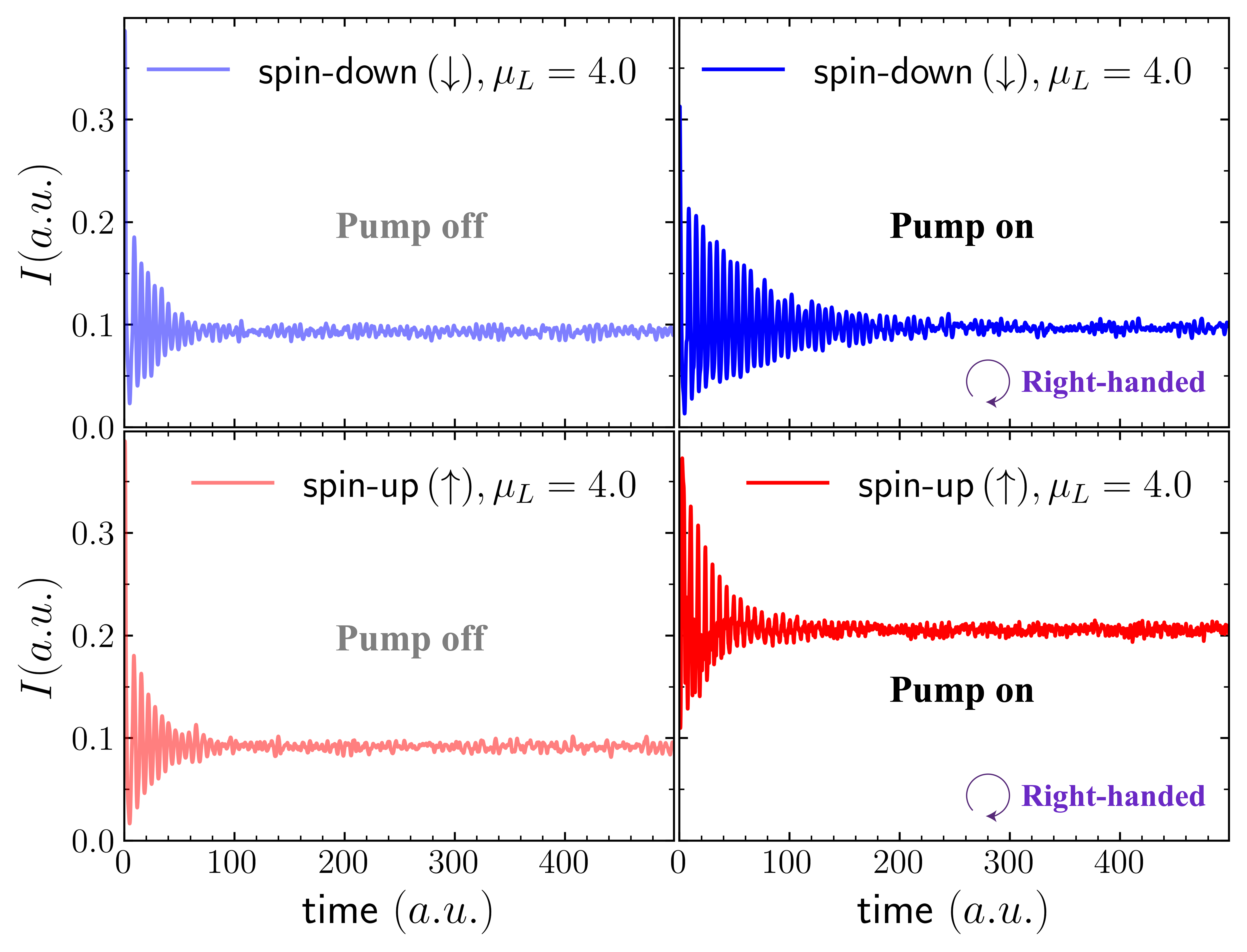}
\caption{Dynamics of spin current $I^{\uparrow/\downarrow}$ in Eq.~\ref{eqn:iupdown_m} for $h_{\text{s}}$ (pump off) in Eq.~\ref{eqn:hs_soc} and $h_{\text{s}}^{\text{[R]}}$ (pump on and right-handed) in Eq.~\ref{eqn:hs_right} when $\mu_{\text{L}}=4.0$ in Fig.~\ref{fig:schematic} (b). All parameters are identical with Fig.~\ref{fig:schematic} (b).}
\label{fig:I_t}
\end{figure}

\textit{Spin Polarization Enhanced by CP Light.}—We present the impact of CP light on enhancing spin polarization. In Fig.~\ref{fig:schematic} (b), we illustrate the steady-state spin current and spin polarization for $\Delta=3$ with $\mu_{\text{L}} = -\mu_{\text{R}}$.

\begin{figure}[htbp]
\includegraphics[width=.48\textwidth]{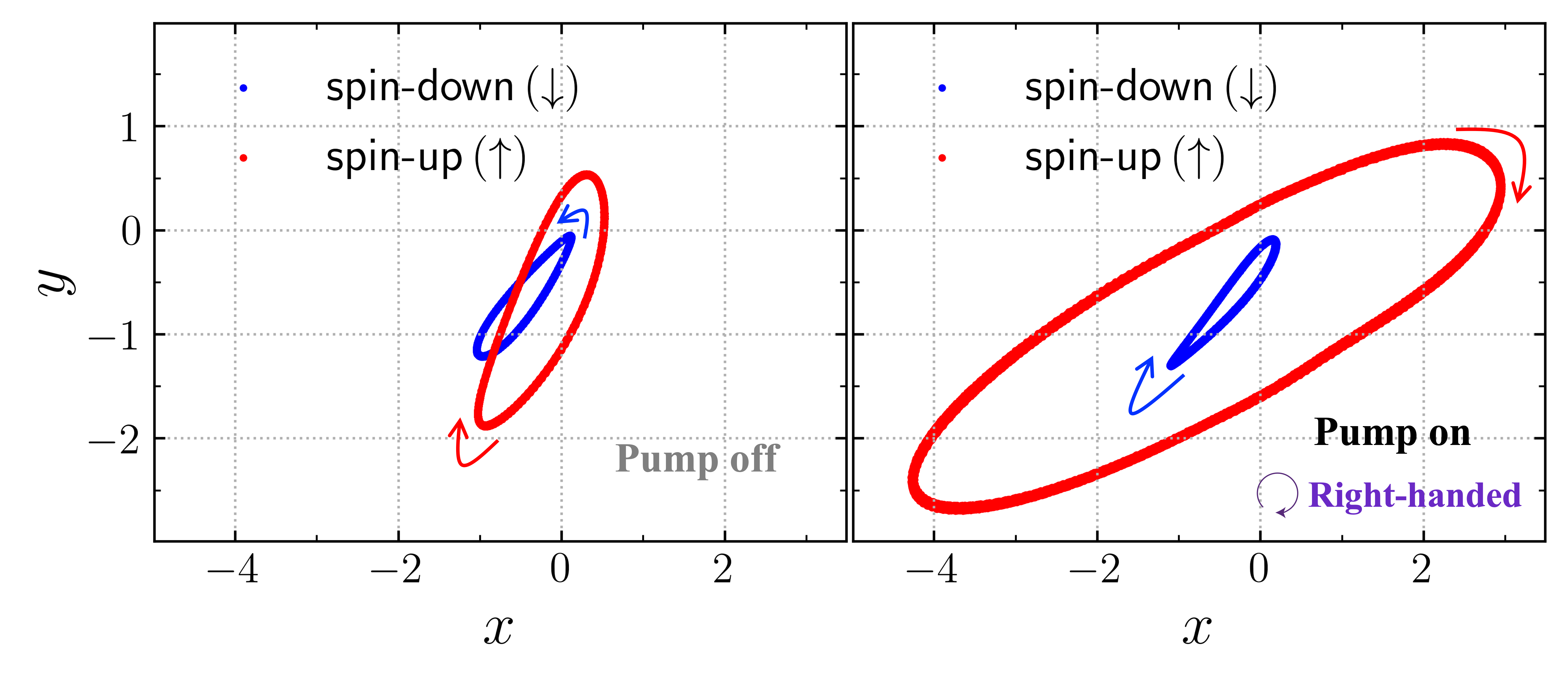}
\caption{Nuclear distribution $\rho^{\uparrow/\downarrow}$ ($300 \leq \text{time} \leq 2000$, $\Delta t=0.05$) in Eq.~\ref{eqn:iupdown_m} for $h_s$ (pump off) in Eq.~\ref{eqn:hs_soc} and $h_s^{\text{[R]}}$ (pump on and right-handed) in Eq.~\ref{eqn:hs_right} when $\mu_{\text{L}}=4.0$ without the random force $\delta F_\mu^{\text{F}}$ from the Langevin dynamics to obtain the steady state distribution. All parameters are identical with Fig.~\ref{fig:I_t}.}
\label{fig:rotation}
\end{figure}

In the left panel (pump off) of Fig.~\ref{fig:schematic} (b), we plot the spin polarization/spin current without light-matter interactions (using $h_s$ in Eq.~\ref{eqn:hs_soc}). Without voltage bias, there is no spin polarization being observed. This finding agrees with experiments~\cite{naaman2020chiral, teh2022spin}. When $-4\leq\mu_{\text{L}}<0$, CISS-induced spin polarization consistently remains below $25\%$, whereas for $0<\mu_{\text{L}}\leq4$, the spin polarization induced by CISS nearly diminishes. Notably, in the middle panel (pump on, right-handed) and right panel (pump on, left-handed) of Fig.~\ref{fig:schematic} (b), activating right-handed ($h_s^{\text{[R]}}$) or left-handed ($h_s^{\text{[L]}}$) CP light results in high spin polarization ($|\xi|\approx 91\%$) at a small voltage bias ($\mu_{\text{L}} = \pm 0.5$). This spin polarization gradually diminishes as $|\mu_{\text{L}}|$ increases, consistently remaining above $25\%$ within the range of $0<|\mu_{\text{L}}|\leq 4$. The symmetric distribution of spin polarization caused by right-handed and left-handed CP light with respect to $0\%$ is consistent with experiments~\cite{gohler2011spin}. Therefore, for simplicity, we only present the results of the right-hand CP light in Fig.~\ref{fig:I_t} and Fig.~\ref{fig:rotation}. Note that the presence of $\Delta=3$ results in an asymmetric energy distribution for the two spatial orbitals. Therefore, the spin polarization for $\mu_{\text{L}}>0$ decreases more rapidly compared to $\mu_{\text{L}}<0$. Within the current range of results, it is evident that the decline in spin polarization persists with the increasing magnitude of $|\mu_{\text{L}}|$. Hence, we recommend employing a small $V_{\text{bias}}$ for researchers aiming to observe conspicuous spin polarization in experiments. 

In Fig.~\ref{fig:I_t}, we plot the dynamics of spin current with the comparison between no light (pump off) and right-handed CP light (pump on) when $\mu_{\text{L}} = 4.0$ in Fig.~\ref{fig:schematic} (b). Spin current fluctuations arise from random forces and progressively diminish with an increasing number of trajectories, set at $10^4$ for each spin state in this calculation. For spin-down electrons, CP light does not result in a significant enhancement of the steady-state current value; however, the time required to reach the steady state considerably prolongs, as illustrated in the upper panel of Fig.~\ref{fig:I_t}. In contrast, for spin-up electrons, the activation of CP light enhances both the steady-state value and the relaxation time of the spin current, as depicted in the lower panel of Fig.~\ref{fig:I_t}. This observation suggests that in this system, CP light introduces additional oscillations, and right-handed CP light specifically exhibits a more pronounced selectivity toward spin-up electrons.

In Fig.~\ref{fig:rotation}, we illustrate the nuclear distribution at $\mu_{\text{L}}=4.0$ without the random force from Langevin dynamics, allowing us to obtain the steady-state distribution~\cite{bode2011scattering}. The distinct steady states observed in nuclei with spin-up and spin-down electrons, as depicted in Fig.~\ref{fig:rotation}, can be ascribed to the influence of the Lorentz-like force associated with the anti-symmetric friction tensor $\gamma_{xy}^{\text{A}}$~\cite{mosallanejad2023floquet}. When pump off, the steady state of nuclear motion exhibits counterclockwise rotation for spin-down electrons and clockwise rotation for spin-up electrons, as illustrated in Fig.\ref{fig:rotation} (pump off). At this stage, the difference in the distribution range between spin-up and spin-down states is minimal, as reflected in the spin current, where the distinction remains inconspicuous (Fig.\ref{fig:I_t}, pump off). Notably, upon the activation of CP light, the contrast in nuclear steady-state distribution between spin-up and spin-down electrons undergoes a significant amplification, as illustrated in Fig.~\ref{fig:rotation} (pump on). This result corresponds with the marked rise in spin polarization depicted in Fig.~\ref{fig:schematic} (b) [$\mu_{\text{L}}=4.0$ for pump off and on (right-handed)]. The introduction of CP light emerges as a crucial factor in enhancing the distinctions in nuclear motion and spin polarization between spin-up and spin-down electronic states.

These findings highlight the critical role of the CP light in influencing the steady states of nuclear motion and, consequently, spin polarization in electronic systems. 
In addition to the specific cases discussed here, our results have broader implications for the fields of spintronics and optoelectronics. The ability to manipulate spin polarization through light-matter interactions opens up the possibilities for developing spin-based devices with tunable properties. Furthermore, the understanding of the underlying mechanisms, such as the influence of CP light, provides guidance for spin control and manipulation in designing novel materials and structures.

\textit{Conclusions.}—In conclusion, our investigation into the influence of CP light on spin polarization in electronic systems has yielded valuable insights. We have demonstrated that CP light can exert a profound enhancement ($>90\%$) on spin polarization. In particular, light-matter interactions introduce Lorentz-like force that changes the nuclear dynamics dramatically for different spin states. As a result, even small spin-orbit couplings (SOC) can introduce large spin selectivity. We find that the interplay of the light-matter interactions and SOC can enhance the spin polarization in quantum transport. 

These findings not only advance our fundamental understanding of spin dynamics in optoelectronic systems but also open up exciting opportunities for practical applications in the realm of spintronics. The ability to control and manipulate spin polarization through light-matter interactions holds great promise for the development of novel technologies, including spin-based quantum computing and information storage.

As we peer into the future, several promising avenues for further research emerge. In the realm of electronic structure theory, there is an urgent need for high-precision calculations of excited-state electronic spin states in large systems, surpassing the capabilities of traditional Density Functional Theory (DFT). Accurately modeling spin-orbit coupling (SOC) as well as derivative couplings is equally crucial. These works are on-going.  

This material is based upon work supported by National Natural Science Foundation of China (NSFC No. 22361142829). W.L. acknowledges support from the high-performance computing center of Westlake University.

\bibliography{Main_ref}
\nocite{*}

\end{document}